\documentclass[aps,pra,twocolumn,10pt,superscriptaddress,longbibliography,nofootinbib]{revtex4-2}

\usepackage{amsmath,amssymb,amsthm,amsfonts}
\usepackage{mathtools}
\usepackage{latexsym}

\usepackage{graphicx}

\usepackage{times,txfonts}
\usepackage{braket}
\usepackage{physics}

\usepackage{blkarray}
\usepackage{tabularx,booktabs}
\usepackage{multirow}

\usepackage{comment}
\usepackage{pifont}
\usepackage[table]{xcolor}
\usepackage{soul}
\usepackage{float}
\usepackage{array}

\usepackage[colorlinks=true,citecolor=blue,urlcolor=blue]{hyperref}
\usepackage{cleveref}

\newcommand{\ba}{\[\begin{aligned}}
\newcommand{\ea}{\end{aligned}\]}

\newcommand{\RNum}[1]{\uppercase\expandafter{\romannumeral #1\relax}}

\newcommand{\be}{\begin{equation}}
\newcommand{\ee}{\end{equation}}

\begin{document}

\title{Beyond-quantum correlations for single system in pseudo-quantum model}
\author{Vishnu Purushothaman}
\email{vishnu_purushothaman@icloud.com}
\affiliation{Department of Physics, Indian Institute of Technology Hyderabad, Kandi, Sangareddy, Telengana-502284, India}
\author{Akarshit Baranwal}
\email{akarshitbaranwal@gmail.com}
\affiliation{Department of Physics, Indian Institute of Technology Hyderabad, Kandi, Sangareddy, Telengana-502284, India}

\author{Rohit Sarma Sarkar }
\email{rohit15sarkar@yahoo.com}
\affiliation{Universitat Politècnica de Catalunya (UPC), Barcelona, Spain-08034}

\author{A.K. Pan }
\email{akp@phy.iith.ac.in}
\affiliation{Department of Physics, Indian Institute of Technology Hyderabad, Kandi, Sangareddy, Telengana-502284, India}
	
	\begin{abstract}
     Within the standard prepare-measure framework, it has been demonstrated that certain compositions of two elementary quantum systems can produce beyond-quantum correlations and can even lead to perfect discrimination of two non-orthogonal quantum states. In this Letter, we introduce the pseudo-quantum model (PQM) for a single system that incorporates the \emph{pseudo-states (effects)} (they satisfy all the properties of quantum states and effects except having negative eigenvalues), and yields valid probability. We consider a prepare-measure communication game known as random access code (RAC) and demonstrate that the success probability achievable within PQM exceeds the corresponding optimal quantum success probability, thereby giving rise to beyond-quantum correlations even for a single system. We then show that such correlations are unphysical as they violate the information causality (IC) principle. Moreover, we demonstrate that two non-orthogonal states can be perfectly distinguishable in PQM for a single system. Therefore, we demonstrate that for a single indivisible system (devoid of any composition), the negativity of \emph{pseudo-states (effects)} can lead to the emergence of beyond-quantum correlations.

	\end{abstract}

	\maketitle

\section{Introduction}
Quantum theory is the most successful theory for describing natural phenomena. The strength of quantum theory lies in the empirical fact that, to date, no experimental observations have been reported that contradict its predictions. The mathematical structure of quantum theory is defined in the Hilbert space, where the states and effects are positive semidefinite matrices. It is postulated that the quantum composition of two elementary quantum systems are so restricted that the state space is self-dual to the effect space \cite{muller_selfduality, wilce_selfduality, janotta14}. Although the physical and conceptual justification for postulating this self-dual structure remains under active discussion, it is noteworthy that so far no experimental evidence has been found that violates this assumption.  Nevertheless, the potential existence of alternative forms of composition replicating the quantum statistics cannot be dismissed. 

In pursuit of a deeper understanding of quantum theory, the general probabilistic theory (GPT) framework is developed \cite{janotta14, barnum06, barnum10march, janotta11, muller11, plavala23, namioka69, arai_aqt, Lami, ak_pan_context_GPT, pan25, hayashi23, chiribella_NRH, spekken_NRH, Janotta_NRH, pan26}. In GPT, the states and effects of an elementary quantum system defined in the Hilbert space can be equivalently represented by positive semi-definite matrices defined in a closed convex cone in the real vector space. Within this framework, a bipartite system may be composed in various possible ways. However, the principles of no-signaling and local tomography single out two extreme compositions known as the minimal and the maximal composition \cite{namioka69}. Any other form of composition, including quantum composition, is placed within these two extreme compositions \cite{janotta14, chiribella_NRH, spekken_NRH, Janotta_NRH,hayashi23, pan26}. 

In a bipartite space-like scenario, it is proven \cite{barnum10} that quantum and beyond-quantum compositions cannot be distinguished if local measurements are quantum. However, this inference does not hold for multipartite scenario \cite{acin10}. In the prepare-measure (time-like) scenario, a no-go theorem called the no-hypersignaling principle \cite{buscemi_NHS, buscemi23} was introduced, that rules out the existence of minimal compositions of two $g$-bits. However, a single $g$-bit itself can produce Popescu-Rohrlich box-like behavior, which is unphysical according to the IC principle \cite{pawloski_IC}.  In \cite{pairwise_distingush_max_comp, pairwise_distingush_min_comp}, a pairwise distinguishability game was proposed that distinguishes extreme compositions from quantum, and in \cite{manik_ic}, it is shown that the IC principle rules out such extreme compositions. In a GPT that allows beyond-quantum composition, it has been demonstrated that the minimum error probability in state discrimination can be reduced \cite{perfect_discrimin_non-orthogonal_state, hayashi_state_discrimin} than quantum theory, and two non-orthogonal quantum states can be perfectly distinguished \cite{perfect_discrimin_non-orthogonal_state}.

Note that by beyond-quantum states and effects in a GPT, we commonly refer that they follow all the properties of quantum states and effects except the positivity. In fact, it is the negativity of the states and effects that can contribute to realizing beyond-quantum correlation \cite{barnum10, acin10, pairwise_distingush_min_comp, pairwise_distingush_max_comp, manik_ic, buscemi_NHS, buscemi23, perfect_discrimin_non-orthogonal_state, hayashi_state_discrimin}. Arai \emph{et al.} \cite{hayashi_state_discrimin} made a direct connection with the success probability of the state discrimination game and the eigenvalue of negative effects in a beyond-quantum composition. The authors identified that, the transition from negativity to positivity is the key factor in deriving quantum theory from beyond-quantum compositions within GPT.

Above results prompted us to introduce the PQM framework for a single system by allowing the sets of beyond-quantum states and effects, yielding a valid probability. We use the terminology \emph{pseudo-states (effects)} to denote the negative-eigenvalued states (effects) in such a PQM. Note that such a PQM can be succinctly mapped in the GPT framework, but this is not the purpose of this work and hence we refrain from doing that. 

In this Letter, we ask the question whether the similar results discussed in \cite{pairwise_distingush_min_comp, pairwise_distingush_max_comp, manik_ic, perfect_discrimin_non-orthogonal_state, hayashi_state_discrimin} can be demonstrated in a PQM of a single indivisible system, having no composition. We answer this question affirmatively. For this, we introduce a prepare-measure communication game in the PQM of a three-level system. Specifically, we consider the $3$-bit prepare-measure RAC, where the sender, Alice, communicates a three-level system to the receiver, Bob. We derive the success probability in classical, quantum, and PQM. We demonstrate that the optimal quantum success probability surpasses its classical counterpart, but the success probability in PQM exceeds the optimal quantum success probability. This thereby showcases the beyond-quantum correlation in PQM featuring a single system. Interestingly, we show that the IC principle is violated, and therefore, such beyond-quantum correlations are unphysical. 

Further, we demonstrate the perfect discrimination of two nonorthogonal states in PQM. All of the above results establish the fact that the presence of negativity in states (or effects) can also be associated with realizing beyond-quantum correlations in our single-system PQM even if there is no composition.

\section{Formulation of PQM }

We take a minimalist approach in defining the PQM for a single system. The state space and effect space in {PQM} is defined as follows.

    The state space  $\Omega$ consists of a set of linear Hermitian operators $\mathcal{L} (\mathcal{H})$, defined in the Hilbert space $\mathcal{H}$. This implies that $\Omega$ includes the density matrices $\rho\in  \mathcal{L} (\mathcal{H}) $ and the \emph{pseudo-state} $\Lambda \in  \mathcal{L} (\mathcal{H})$ which are Hermitian matrices with positivity relaxed.  The state $\Lambda$ preserves unit trace, but the eigenvalue can lie in between $+1$ and $-1$.

   The effect space $\mathcal{E}$ consists of a set of linear Hermitian operators $\mathcal{L} (\mathcal{H})$ including quantum effects $E \in \mathcal{L} (\mathcal{H}) $ and the ones with positivity relaxed are \emph{pseudo-effects} denoted as $\tau \in \mathcal{L} (\mathcal{H})$. The condition of valid outcome probability ensures that only one of the two possible constructions can serve as a valid realization of PQM framework. The former relaxes the state space to accommodate both quantum states and \emph{pseudo-states} while restricting the effect space to a subset of quantum effects. The latter restricts the state space to a subset of quantum states while relaxing the effect space to include both quantum effects and \emph{pseudo-effects}. The state space and effect space form a convex set and are closed under mixing. The condition of valid probability ensures a dual relationship between the state and the effect cone. If all the states and effects are positive semidefinite, the state cone and effect cone become self-dual, and the PQM framework reduces to quantum theory.

In GPT, an elementary quantum system can be represented in the state and effect space defined in the real vector space $\mathbb{R}^{n}$ \cite{janotta14}. For composite systems, the requirements of no-signaling and local tomography restrict the joint state and effect space to lie between the two extremes, commonly referred to as the minimal and maximal composition. In the former, the state space is restricted to product states, while the effect space accommodates quantum and beyond-quantum effects. In the latter, the state space is a maximal set of quantum and beyond-quantum states, while the effect space is restricted to product effects \cite{janotta14, plavala23}. The beyond-quantum states and effects in GPT can equivalently be represented by the corresponding \emph{pseudo-states} and \emph{pseudo-effects} in the PQM.

Between the two extreme compositions in GPT, there lies infinitely many beyond-quantum compositions \cite{Janotta_NRH, janotta11, namioka69, plavala23}. For such intermediate compositions, two possible choices for constructing the state space and effect space are viable. For those compositions containing beyond-quantum state, the corresponding state space is characterized by these beyond-quantum state together with some separable and entangled states that yield valid outcome probabilities for every effect in the effect space. The resulting effect space comprises all product effects and a restricted set of entangled effects chosen to ensure the outcome probability remains valid for every state in the state space. Likewise, for compositions containing some beyond-quantum effects, an analogous construction is obtained by interchanging the roles of states and effects \cite{arai_aqt, hayashi23, hayashi_state_discrimin, Janotta_NRH, plavala23, janotta11}. Moreover, by the additional assumption of the no-restriction hypothesis, all those logical probability rules realized by the states and effects are assumed to be experimentally realizable \cite{chiribella_perinotti10, spekken21, janotta14}.

Inspired by the existence of such intermediate compositions that respect valid outcome probabilities, we consider two possible choices for constructing the state and effect space in the PQM model that likewise yield valid probabilities. Each construction represents a valid realization of the single system PQM model.

\section{The IC principle}

The principle, as initially proposed in \cite{pawloski_IC, pawloski_ic_bound}, is known as the generalization of the no-signaling principle and is adhered to by all physical theories. This principle is deduced independently of any specific theory, employing certain information-theoretic properties such as: $(i)$ the Consistency condition, $(ii)$ the Data processing inequality, and $(iii)$ the Chain rule. In its foundational version \cite{pawloski_IC, pawloski_ic_bound}, the IC principle is derived within a two-party RAC framework. Here, Alice obtains a uniformly random $n$-bit sequence, taken from $x \in \{0,1\}^{n}$, while Bob receives the index $y \in \{1,2,\ldots,n\}$ randomly. Bob's objective is to accurately deduce the bit in Alice's string. It asserts that if any local resource is used jointly by both parties, and Alice transmits $m$ $c$-bits ($m \leq n$) of information to Bob, then the total mutual information between the specific bit $x_{k}$ of Alice and Bob's interpreted bit $b_y$, under the condition where $y=k$, is limited to $m$ bits. The expression gives the general requirement for IC to hold,
\begin{align}
   \mathcal{I} = \sum_{k=1}^{n}\, \mathcal{I}(x_{k}:b_y|y=k) \leq m 
\end{align}

where $\mathcal{I}(x_{k}: b_y|y=k) = H(x_k) + H(b_y) - H(x_k,b_y)$ represents the mutual information, and $H(x_k)$ and $H(b_y)$ denote the Shannon entropy of Alice's $k^{\text{th}}$ bit and Bob's decoded bit $b_y$, respectively, and $H(x_k,b_y)$ denotes their joint entropy.

\section{$3\rightarrow log_{2}3$ classical and quantum RAC} 
We consider a communication game in the prepare-measure scenario, known as $3$-bit RAC, which is played between two parties, Alice and Bob. In such a game, Alice receives length-$3$ strings, randomly sampled from $x\in \{0,1\}^3$ \cite{akpan_PORAC, tavakoli_d_dimen_RAC15, tavakoli_d_dimen_RAC25, pawlowski18}. Bob receives, uniformly at random, the numbers $y \in [3]$. Bob's task is to recover $y^{\text{th}}$ bit of Alice's input string $x$, which is the winning condition of the game. In order to assist Bob, Alice transmits a classical trit containing the encoded information to Bob. Throughout this work, we limit our focus to the $3$-level RAC scenario so that the dimension of the message communicated by Alice to Bob is three.

Within the context of an operational theory framework, Alice uses a preparation procedure denoted as $P_{x}$ to encode her input sequence, $x$, in a specific physical state. This encoded state is subsequently transmitted to Bob. After receiving the state, Bob performs a binary outcome measurement $M_{y}$. The result Bob obtains from this measurement is reported as the outcome $b$. The condition that defines a win is when the outcome is $b = x_{y}$. Consequently, the probability of success for the RAC task can be calculated as
	\begin{align}
		\label{SP}
		\mathcal{P}=\dfrac{1}{24}\sum_{y=1}^{3}\sum_{x\in \{0,1\}^{3}}p(b=x_{y}|P_{x},M_{y}).
	\end{align}
In a classical $3$-bit RAC, Alice transmits a classical trit  $m \in \{0, 1, 2\}$ to help Bob and the optimal strategy is the following.  For bit string in the set $\{000, 001, 010, 100\}$ she sends the message $m=0$ and for the set $\{011, 101, 111\}$ she sends $m=1$. She sends $m=2$ for the string $110$.  The maximum average success probability is derived as $(\mathcal{P})_{C} = 19/24\approx 0.792$.

In the quantum version of the three-level $3$-bit RAC, Alice sends a $3$-level quantum system. Let Alice encode her input string $x$ into qutrit density matrix $\rho_{x}$, prepared by a procedure $P_{x}$, and communicates to Bob. Upon receiving input, Bob performs a binary measurement $\{E_{y}^{b}\}$ ( $\sum_{b}E_{y}^{b}=\mathbb{I}_{3}$ and $b=0,1$) to predict the bit $x_y$, where $y\in[3]$.  The success probability can then be written as 
	\begin{equation}
		\label{QSP}
		(\mathcal{P})_{Q}=\dfrac{1}{24}\sum_{y=1}^{3}\sum_{x\in \{0,1\}^{3}}\mathrm{tr}\hspace{2pt}[\rho_{x}E_{y}^{b}].
	\end{equation}
    In \cite{tavakoli_d_dimen_RAC25},  the lower and upper bounds on the average success probability of three-level  $3$-bit RAC were derived using the Seesaw algorithm and with an analytical approach, respectively. They found the lower bound $0.808$ and an upper bound $0.911$.  Here, we attempt a derivation and find a tighter  upper bound on the average success probability as
\begin{align}
    (\mathcal{P})_{Q} \leq   0.859
\end{align}
The detailed derivation of this upper bound is quite lengthy and is deferred to the Appendix \ref{quant_RAC_success_probab}.\\

\section{$3\rightarrow log_{2}3$ RAC in PQM}

    We now derive the average success probability of $3$-bit RAC within the framework of PQM for two different cases.  In \emph{Case-I}, we consider the PQM where the state space contains only a subset of quantum states, but the effect space contains both quantum and \emph{pseudo-effects}. In \emph{Case-II}, we consider a PQM where the state space allows both quantum and \emph{pseudo-states} but the effect space contains only a subset of quantum effects. We demonstrate that in both cases, the average success probabilities exceed the upper bound of the quantum success probability. Interestingly, for both cases, the IC bound is violated, thereby demonstrating the unphysicality of such beyond-quantum correlations. \\

\subsection{Case-\RNum{1}} 

Alice encodes her input strings into eight pure states $\rho_x =|\psi_x\rangle\langle\psi_x|$ as follows.
\begin{equation*}
\rho_{000}=\frac{1}{6}
\begin{pmatrix}
1 & 1 & 2 \\
1 & 1 & 2 \\
2 & 2 & 4 
\end{pmatrix}
\hspace{0.3cm}
\rho_{001}=
\begin{pmatrix}
0 & 0 & 0 \\
0 & 0 & 0 \\
0 & 0 & 1 
\end{pmatrix}
\hspace{0.3cm}
\rho_{010}=\frac{1}{2}
\begin{pmatrix}
1 & 0 & 1\\
0 & 0 & 0 \\
1 & 0 & 1
\end{pmatrix}
\end{equation*}

\begin{equation}
\rho_{011}=\frac{1}{2}
\begin{pmatrix}
1 & 0 & -1\\
0 & 0 & 0 \\
-1 & 0 & 1 
\end{pmatrix}
\hspace{1cm}
\rho_{100}=\frac{1}{2}
\begin{pmatrix}
0 & 0 & 0 \\
0 & 1 & 1 \\
0 & 1 & 1
\end{pmatrix}
\end{equation}

\begin{equation*}
\rho_{101}=\frac{1}{2}
\begin{pmatrix}
0 & 0 & 0 \\
0 & 1 & -1 \\
0 & -1 & 1
\end{pmatrix}
\hspace{0.1cm}
\rho_{110}=\frac{1}{2}
\begin{pmatrix}
1 & 1 & 0 \\
1 & 1 & 0 \\
0 & 0 & 0 
\end{pmatrix}
\hspace{0.1cm}
\rho_{111}=\frac{1}{2}
\begin{pmatrix}
1 & -1 & 0 \\
-1 & 1 & 0 \\
0 & 0 & 0 
\end{pmatrix}
\end{equation*}

Bob performs the measurement of \emph{pseudo-effects}   $\{ \tau_y,\ \overline{\tau} _y\}$ for $y=1$ and $y=2$, to retrieve the first and second bit of Alice's string $x$, respectively. For the third bit, he uses the quantum effect $\{ E_3, \overline{E}_3\}$.  Bob's choices of effects are given by
\begin{align}
\tau_{1}=
\begin{pmatrix}
1 & 0 & 0 \\
0 & -1 & 0 \\
0 & 0 & 1 
\end{pmatrix} \hspace{0.5cm}
\tau_{2}=
\begin{pmatrix}
-1 & 0 & 0 \\
0 & 1 & 0 \\
0 & 0 & 1 
\end{pmatrix}\hspace{0.5cm}
E_{3}=\frac{1}{3}
\begin{pmatrix}
1 & 1 & 1 \\
1 & 1 & 1 \\
1 & 1 & 1 
\end{pmatrix}
\end{align}
Bob decodes the bit as 0 (1) if $\tau_y$ $(\overline{\tau}_y)$ or $E_3$ $(\overline{E}_3)$ clicks, where $\overline{\tau}_y =\mathbb{I} _{3} -\tau_y$ and $\overline{E}_3 =\mathbb{I}_{3} -E_3$.
The probability of decoding each individual bit is given in tabular form in Table \ref{probab_table_case1}, in Appendix \ref{appendix_probab_table}. We derive the average  success probability of this particular RAC task for decoding each bit in \emph{{Case-I}} as

\begin{align}
\label{ss1}
    (\mathcal{P})^{I}_{PQM}= 0.912
\end{align}

which exceeds the upper bound in quantum theory. We also note that a better choice of states and effects may be found, which can provide a higher success probability than in Eq. (\ref{ss1}).

We show that the IC principle is violated by the beyond-quantum correlation that arises in PQM.  Using Table \ref{probab_table_case1}, in Appendix \ref{appendix_probab_table}, we calculate $\mathcal{I} \approx 1.916 > log_{2}3$.  Therefore, IC principle is violated for the beyond-quantum correlation arising in PQM. \\

\subsection{Case-\RNum{2}}

In contrast to \emph{Case-I}, here  Alice encodes the $3$-bit string into \emph{pseudo-states} and pure quantum states.  The encoding scheme chosen by Alice is the following. 
\begin{equation*}
\Lambda_{000}= \frac{1}{3}
\begin{pmatrix}
2 & 2 & -1 \\
2 & -1 & 2 \\
-1 & 2 & 2
\end{pmatrix}
\hspace{0.1cm}
\Lambda_{001}= 
\begin{pmatrix}
1 & 0 & 0 \\
0 & -1 & 0 \\
0 & 0 & 1 
\end{pmatrix}
\hspace{0.1cm}
\Lambda_{010}=
\begin{pmatrix}
1 & 0 & 0 \\
0 & 0 & 1 \\
0 & 1 & 0 
\end{pmatrix}
\end{equation*}

\begin{equation*}
\rho_{011}=
\begin{pmatrix}
1 & 0 & 0 \\
0 & 0 & 0 \\
0 & 0 & 0 
\end{pmatrix}
\hspace{0.5cm}
\Lambda_{100}=
\begin{pmatrix}
0 & 1 & 0 \\
1 & 0 & 0 \\
0 & 0 & 1 
\end{pmatrix}
\hspace{0.5cm}
\rho_{101}=
\begin{pmatrix}
0 & 0 & 0 \\
0 & 0 & 0 \\
0 & 0 & 1 
\end{pmatrix}
\end{equation*}

\begin{align}
\Lambda_{110}=
\begin{pmatrix}
0 & 0 & 1 \\
0 & 1 & 0 \\
1 & 0 & 0 
\end{pmatrix}
\hspace{0.5cm}
\rho_{111}=
\begin{pmatrix}
0 & 0 & 0 \\
0 & 1 & 0 \\
0 & 0 & 0 
\end{pmatrix}
\end{align}

where $\Lambda_x$ denotes the  \emph{pseudo-state}, i.e., the beyond-quantum state. 

Bob performs the measurement of quantum effects $\{ E_y, \overline{E}_y=\mathbb{I}_{3}-E_y\}$ for  where $y=1,2$ and $3$ to decode each individual bit. Bob's choice of effects is given by
\begin{align}
{E_{1}}=
\begin{pmatrix}
1 & 0 & 0 \\
0 & 0 & 0 \\
0 & 0 & 0 
\end{pmatrix} \hspace{0.5cm}
{E_{2}}=
\begin{pmatrix}
0 & 0 & 0 \\
0 & 0 & 0 \\
0 & 0 & 1 
\end{pmatrix}\hspace{0.5cm}
{E_{3}}=\frac{1}{3}
\begin{pmatrix}
1 & 1 & 1 \\
1 & 1 & 1 \\
1 & 1 & 1 
\end{pmatrix}
\end{align}
The calculated probabilities for each bit are placed in Table \ref{probab_table_case2} in Appendix \ref{appendix_probab_table}. In \emph{Case-II}, we derive the average success probability,

\begin{align}
    (\mathcal{P})^{\RNum{2}}_{PQM} = 0.916
\end{align}

which again exceeds the quantum upper bound. Using Table \ref{probab_table_case2} in Appendix \ref{appendix_probab_table} we derive the IC bound $\mathcal{I} \approx 2.035 > \log_{2}3$.  

Therefore, in the proposed PQM framework, we showed two scenarios where the \emph{pseudo-states (effects)} give advantage in achieving beyond-quantum success probability for a $3$-bit RAC. The resulting beyond-quantum correlations in both scenarios are unphysical as they violate the IC principle. 

We note that similar results were demonstrated in \cite{manik_ic} where they argued that the IC principle rules out extreme compositions of two elementary quantum systems. We consider a single system PQM having no possibility of composition and demonstrate that the IC principle invalidates the beyond-quantum correlation arising in PQM. We argue that it is the negativity of the states or effects that is the origin of enabling beyond-quantum correlation.  This will be further established in the following state discrimination task.

\section{Distinguishability of two quantum states}

In quantum theory, two non-orthogonal states cannot be perfectly distinguished.  Arai \emph{et al.,} \cite{perfect_discrimin_non-orthogonal_state} demonstrated the perfect discrimination of two non-orthogonal states in a GPT that allows beyond-quantum composition of two elementary qubit systems. They explicitly made a connection between the negativity of effects and the probability of state discrimination. They showed that the state discrimination probability corresponds to quantum theory when the effect becomes positive \cite{hayashi_state_discrimin}.

We examine whether two non-orthogonal qutrit states can be perfectly distinguished in PQM. We consider the states as
\begin{equation}
\rho_{1}=
\begin{pmatrix}
1 & 0 & 0 \\
0 & 0 & 0 \\
0 & 0 & 0 
\end{pmatrix}
;\hspace{0.5cm}
\rho_{2}= \frac{1}{2}
\begin{pmatrix}
1 & 1 & 0 \\
1 & 1 & 0 \\
0 & 0 & 0 
\end{pmatrix}
\hspace{0.5cm}
\end{equation}
which has a overlap of $1/2$. In quantum theory, these states cannot be perfectly distinguished. 

We consider the \emph{pseudo-effects} $\{\tau_1, \tau_2 \}$ where 
\begin{equation}
\tau_{1}=
\begin{pmatrix}
1 & -1 & 0 \\
-1 & 1 & 0 \\
0 & 0 & 0 
\end{pmatrix}
\hspace{0.5cm}
\tau_{2}= 
\begin{pmatrix}
0 & 1 & 0 \\
1 & 0 & 0 \\
0 & 0 & 1 
\end{pmatrix}
\hspace{0.5cm}
\end{equation}
with $\tau_{1}+\tau_{2}=\mathbb{I}_{3}$. Interestingly, the $\{\tau_1, \tau_2 \}$ perfectly distinguishes the non-orthogonal states $\rho_1$ and $\rho_2$.

\section{Summary and discussion}  

In sum, we demonstrated that beyond-quantum correlation can arise in the PQM, featuring a single indivisible system. Based on a communication game known as RAC we demonstrated that the average success probability in the PQM surpasses the optimal quantum success probability. Further, we demonstrated that two non-orthogonal states can be perfectly distinguished in PQM. We also showed that such a beyond-quantum correlation arising in PQM violates the IC principle; therefore, they are unphysical. The above findings resemble the results already obtained in \cite{pairwise_distingush_min_comp, pairwise_distingush_max_comp, manik_ic, perfect_discrimin_non-orthogonal_state, hayashi_state_discrimin} but arising from the beyond-quantum composition of two elementary quantum systems.  

The notion of the composition of two elementary quantum systems in a GPT has been widely studied. While space-like scenario cannot distinguishes the quantum and beyond-quantum compositions, in time-like scenario, there are numerous results that demonstrate beyond-quantum correlations. It is believed that such correlations arise in a GPT that allows beyond-quantum compositions. Note that all the above results in \cite{pairwise_distingush_min_comp, pairwise_distingush_max_comp, manik_ic, perfect_discrimin_non-orthogonal_state, hayashi_state_discrimin} uses \emph{pseudo-states (effects)} having negative eigenvalues. Arai \emph{et al.,} \cite{hayashi_state_discrimin} have shown a direct connection between state distinguishability and negativity of effects in beyond-quantum compositions of two systems.

In contrast, we demonstrated that beyond-quantum correlation can arise in single system if there exists a super theory (we call it as PQM) allowing \emph{pseudo-states (effects)}. Therefore, beyond-quantum correlations not only originate from the composition of two elementary quantum systems, but also from the negativity of the eigenvalues of the states (or effects).  Consequently, what the IC principle actually rules out is the \emph{pseudo-states (effects)} rather than specific composition of two elementary quantum systems. 

We note again that by relaxing positivity, PQM  captures beyond-quantum correlation for a single system devoid of composition. However, it has been shown \cite{acin_almost_quantum} that by relaxing the tensor product structure of composite systems to dot product in the global Hilbert space (following a strict commutation relation), the post-quantum correlations can also be realized. Hence in a broader GPT framework, there can be many ways to realize beyond-quantum correlations and PQM is one among them.

\appendix

\section{Derivation of quantum success probability} \label{quant_RAC_success_probab}
By writing $E^{b}_{y}=\frac{\mathbb{I} _{3} \pm B_{y}}{2}$ where $y=1,2,3$, the success probability $(\mathcal{P})_{Q}$ in Eq. (\ref{QSP}) can explicitly be written as

\begin{align}
\label{dd}
\nonumber
(\mathcal{P})_Q=\frac{1}{2} + &\frac{1}{48}\mathrm{tr}\hspace{2pt}\Big[
     \big(\rho_{000} +\rho_{011} -\rho_{101}-\rho_{110} \\
     &+ \rho_{001} +\rho_{010} -\rho_{100}-\rho_{111}\big)B_1 \nonumber\\
      &+\big(\rho_{000} -\rho_{011} +\rho_{101}- \rho_{110}\\
      &
      \nonumber +\rho_{001} -\rho_{010} +\rho_{100}-\rho_{111}\big)B_2\\
      &\nonumber+\big(\rho_{000} -\rho_{011} -\rho_{101}+\rho_{110}\\
      & \nonumber
      -\rho_{001} +\rho_{010} +\rho_{100}- \rho_{111}\big)B_3 \Big]
\end{align}

We introduce observables $B_1 = \rho_{000}+\rho_{011}-\rho_{111}$,\, $B_2 = \rho_{001}+\rho_{101}-\rho_{010}$\, and \,$B_3 = \rho_{110}+\rho_{100}-\rho_{111}$ so that   {$\rho_{000}+\rho_{011}+\rho_{111}= \rho_{001}+\rho_{101}+\rho_{010}= \rho_{110}+\rho_{100}+\rho_{111}=\mathbb{I}_{3}$. They all are eigenstates and hence pure states. Putting those to Eq. (\ref{dd}) we have

\begin{align}
\label{dd11}
\nonumber
(\mathcal{P})_Q\leq\frac{1}{2} + &\frac{1}{48}\Bigg( 9 + \mathrm{tr}\hspace{2pt} \Big[
     \big( -\rho_{101}-\rho_{110}+\rho_{001}+\rho_{010}-\rho_{100}\big)B_1 \\
     &
     \nonumber
      +\big(\rho_{000}-\rho_{011}-\rho_{110}+\rho_{100}-\rho_{111}\big)B_2\nonumber\\
      &+\big(\rho_{000}-\rho_{011}-\rho_{101}-\rho_{001}+\rho_{010}\big)B_3 \Big] \Bigg)
\end{align}
Now, assuming $ \omega_1= ||-\rho_{101}-\rho_{110}+\rho_{001}+\rho_{010}-\rho_{100}||$, $\omega_2 = ||\rho_{000}-\rho_{011}-\rho_{110}+\rho_{100}-\rho_{111}||$ and $ \omega_3= ||\rho_{000}-\rho_{011}-\rho_{101}-\rho_{001}+\rho_{010}||$, we can further re-write Eq. (\ref{dd11}) as
\begin{align}
\label{dd2}
(\mathcal{P})_Q\leq\frac{1}{2} + &\frac{1}{48}\Big( 9+ \mathrm{tr}\hspace{2pt} \big[ \omega_1 +\omega_2 +\omega_3
      \big] \Big)
\end{align}
where 
\begin{align}
\label{dd5}
\nonumber
    \omega_1
    &      = \big[5 + \mathrm{tr}[\{\rho_{101},\mathbb{I}_{3}\}-\{\rho_{101},\rho_{111}\}-\{\rho_{110},\rho_{001}\}\\
    &
    \nonumber-\{\rho_{110},\rho_{010}\}-\{\rho_{001},\rho_{100}\}-\{\rho_{010},\rho_{100}\}]\big]^{\frac{1}{2}}\\
         \omega_2& = 
      \big[5+\mathrm{tr}[-\{\rho_{000},\mathbb{I}_{3}\}+\{\rho_{000},\rho_{100}\}\\
    &
    \nonumber+\{\rho_{000},\rho_{100}\}+\{\rho_{011},\rho_{110}\}-\{\rho_{011},\rho_{100}\}]\big]^{\frac{1}{2}}\\
      \nonumber
       \omega_3& = \big[5+\mathrm{tr}[\rho_{010}^{2}-\{\rho_{000},\rho_{101}\}-\{\rho_{000},\rho_{001}\}+\{\rho_{000},\rho_{010}\}\\
    &
    \nonumber+\{\rho_{011},\mathrm{I}_{3}\}-\{\rho_{011},\rho_{010}\}-\{\rho_{011},\rho_{010}\}\big]^{\frac{1}{2}}
    \end{align}
From Eq. (\ref{dd5}), we finally get $ \max(\omega_1+\omega_2+\omega_3)
= \sqrt{5}+6$
The success probability is then
\begin{align}
(\mathcal{P})_Q\leq\frac{1}{2} + \frac{15+\sqrt{5}}{48}\approx 0.859
\end{align}

 \section{Derived probabilities in \emph{Case-\RNum{1}} and \emph{Case-\RNum{2}} in PQM} \label{appendix_probab_table} 

% \begin{table}[h]
%     \centering
% \begin{tabular}{|p{2.1cm}|p{1.9cm}|p{1.9cm}|p{1.9cm}|}
%  \hline
% Alice Encoding  & \multicolumn{3}{c|}{Bob's success probability for guessing each bit.} \\
%  \cline{2-4}
%  \hspace{0.5cm}$x \in \{0,1\}^{3}$ & \hspace{0.2cm} $1^{st}$ bit& \hspace{0.2cm} $2^{nd}$ bit & \hspace{0.2cm} $3^{rd}$ bit\\
%  \hline
%  \hline
%   \hspace{0.8cm}$\rho_{000}$   & \hspace{0.5cm}  2/3  &  \hspace{0.5cm} 2/3  & \hspace{0.6cm}   8/9 \\
%  \hline
% \hspace{0.8cm}$\rho_{001}$  & \hspace{0.6cm} 1   & \hspace{0.6cm} 1   & \hspace{0.6cm} 2/3 \\
% \hline
% \hspace{0.8cm}$\rho_{010}$  &  \hspace{0.6cm} 1 & \hspace{0.6cm} 1  &  \hspace{0.6cm}   2/3\\
% \hline
% \hspace{0.8cm}$\rho_{011}$   &  \hspace{0.6cm} 1  & \hspace{0.6cm} 1    & \hspace{0.7cm} 1\\
%  \hline
% \hspace{0.8cm}$\rho_{100}$  & \hspace{0.6cm} 1    & \hspace{0.6cm} 1  & \hspace{0.6cm} 2/3 \\
% \hline
% \hspace{0.8cm}$\rho_{101}$  & \hspace{0.6cm} 1   & \hspace{0.6cm} 1  & \hspace{0.7cm} 1 \\
% \hline
% \hspace{0.8cm}$\rho_{110}$   & \hspace{0.6cm} 1  & \hspace{0.6cm} 1  & \hspace{0.6cm} 2/3  \\
%  \hline
% \hspace{0.8cm}$\rho_{111}$ & \hspace{0.6cm} 1  & \hspace{0.6cm} 1  & \hspace{0.7cm} 1\\
%  \hline
% \end{tabular}
% \caption{Success probabilities in \emph{Case-\RNum{1}}.}
% \label{probab_table_case1} 
% \end{table}

\begin{table}[h]
    \centering
    \begin{tabular}{|c|c|c|c|}
        \hline
        Alice Encoding & \multicolumn{3}{c|}{\begin{tabular}{@{}c@{}}Bob's success probability\\for guessing each bit.\end{tabular}} \\
        \cline{2-4}
        $x \in \{0,1\}^{3}$ & \makebox[1.65cm][c]{$1^{st}$ bit} & \makebox[1.65cm][c]{$2^{nd}$ bit} & \makebox[1.65cm][c]{$3^{rd}$ bit} \\
        \hline
        \hline
        $\rho_{000}$ & $2/3$ & $2/3$ & $8/9$ \\
        \hline
        $\rho_{001}$ & $1$ & $1$ & $2/3$ \\
        \hline
        $\rho_{010}$ & $1$ & $1$ & $2/3$ \\
        \hline
        $\rho_{011}$ & $1$ & $1$ & $1$ \\
        \hline
        $\rho_{100}$ & $1$ & $1$ & $2/3$ \\
        \hline
        $\rho_{101}$ & $1$ & $1$ & $1$ \\
        \hline
        $\rho_{110}$ & $1$ & $1$ & $2/3$ \\
        \hline
        $\rho_{111}$ & $1$ & $1$ & $1$ \\
        \hline
    \end{tabular}
    \caption{Success probabilities in \emph{Case-\RNum{1}}.}
    \label{probab table case1}
\end{table}

\begin{table}[ht]
    \centering
    \begin{tabular}{|c|c|c|c|}
        \hline
        Alice Encoding & \multicolumn{3}{c|}{\begin{tabular}{@{}c@{}}Bob's success probability\\for guessing each bit\end{tabular}} \\
        \cline{2-4}
        $x \in \{0,1\}^{3}$ & \makebox[1.65cm][c]{1st bit} & \makebox[1.65cm][c]{2nd bit} & \makebox[1.65cm][c]{3rd bit} \\
        \hline
        $\Lambda_{000}$ & $2/3$ & $2/3$ & $1$ \\
        \hline
        $\Lambda_{001}$ & $1$ & $1$ & $2/3$ \\
        \hline
        $\Lambda_{010}$ & $1$ & $1$ & $1$ \\
        \hline
        $\rho_{011}$ & $1$ & $1$ & $2/3$ \\
        \hline
        $\Lambda_{100}$ & $1$ & $1$ & $1$ \\
        \hline
        $\rho_{101}$ & $1$ & $1$ & $2/3$ \\
        \hline
        $\Lambda_{110}$ & $1$ & $1$ & $1$ \\
        \hline
        $\rho_{111}$ & $1$ & $1$ & $2/3$ \\
        \hline
    \end{tabular}
    \caption{Success probabilities in \emph{Case-\RNum{2}}.}
    \label{probab_table_case2}
\end{table}

\emph{Acknowledgments:-}
A.B. acknowledges funding from the DST-Inspire fellowship No. DST/INSPIRE
Fellowship/2023/IF210650, Govt. of India. A.K.P. acknowledges the support from the research grant I-HUB/PDF/2022-23/06, Government of India.

\end{document}